\documentclass[9pt,sigconf,letterpaper]{acmart}

\usepackage[english]{babel}
\usepackage{blindtext}
\usepackage{xcolor}
\usepackage{array}
\usepackage{balance}
\newcolumntype{P}[1]{>{\centering\arraybackslash}p{#1}}

\copyrightyear{2020}
\acmYear{2020}
\setcopyright{acmlicensed}\acmConference[CoNEXT '20]{The 16th International Conference on emerging Networking EXperiments and Technologies}{December 1--4, 2020}{Barcelona, Spain}
\acmBooktitle{The 16th International Conference on emerging Networking EXperiments and Technologies (CoNEXT '20), December 1--4, 2020, Barcelona, Spain}
\acmPrice{15.00}
\acmDOI{10.1145/3386367.3431309}
\acmISBN{978-1-4503-7948-9/20/12}

\begin{document}
\title[Testing Compilers for Prog. Switches Through Switch Hardware Sim.]{Testing Compilers for Programmable Switches Through Switch Hardware Simulation}

\author{Michael D. Wong, Aatish Kishan Varma, Anirudh Sivaraman}

\affiliation{\textit{New York University}}

\def\compactify{\itemsep=0pt \topsep=0pt \partopsep=0pt \parsep=0pt}
\let\latexusecounter=\usecounter
\newenvironment{CompactItemize}
  {\def\usecounter{\compactify\latexusecounter}
   \begin{itemize}}
  {\end{itemize}\let\usecounter=\latexusecounter}
\newenvironment{CompactEnumerate}
  {\def\usecounter{\compactify\latexusecounter}
   \begin{enumerate}}
  {\end{enumerate}\let\usecounter=\latexusecounter}

\newcommand{\eg}{e.g.,\ }
\newcommand{\ie}{i.e.,\ }
\newcommand{\etc}{etc.\xspace}
\newenvironment{parafont}{\fontfamily{ptm}\selectfont}{}
\newcommand{\Para}[1]{\vspace{2pt}\noindent\begin{parafont}\textbf{\textit{#1}}\end{parafont}}

\begin{abstract}

Programmable switches have emerged as powerful and flexible alternatives to fixed-function forwarding devices. But because of the unique hardware constraints of network switches, the design and implementation of compilers targeting these devices is tedious and error prone. Despite the important role that compilers play in software development, there is a dearth of tools for testing compilers for programmable network devices. We present Druzhba, a programmable switch simulator used for testing compilers targeting programmable packet-processing substrates. We show that we can model the low-level behavior of a switch's programmable hardware. We further show how our machine model can be used by compiler developers to target Druzhba as a compiler backend. Generated machine code programs are fed into Druzhba and tested using a fuzzing-based approach that allows compiler developers to test the correctness of their compilers. Using a program-synthesis-based compiler as a case study, we demonstrate how Druzhba has been successful in testing compiler-generated machine code for our simulated switch pipeline instruction set. 
    
\end{abstract}

\ccsdesc[500]{Networks~Programmable networks}
\ccsdesc[500]{Networks~Network simulations}
\keywords{Programmable data plane; simulation; programmable switches; compilers; software testing}
\maketitle

\section{Introduction}

Traditionally, network switches have been \textit{fixed-function}; in these switches, behavior is baked into the underlying hardware itself with little to no room for modification in the field. Though there have been programmable network processors available (\eg \cite{ixp}), it was widely believed that fixed-function switches would always be cheaper, more power efficient, and much faster. Programmable packet-processing devices were failing to reach the 1 Tb/s packet forwarding speeds observed in large data centers and enterprises, causing many to opt not to deploy these systems. However, network operators need to be able to dynamically add new functionalities and packet-processing operations whilst still ensuring that the device runs at high-speeds. While operators can opt to make an investment in new fixed-function hardware, this is a time-consuming and costly use of resources. It can take several years for network switch vendors to produce these new devices due to the complications of designing new software and ASIC hardware. After these switches are developed, it takes additional time and effort to integrate these devices within their networking infrastructure.

The emerging prominence of software-defined networking (SDN) \cite{sdn} and programmable networks has attempted to mitigate these issues. Programmability of the data plane has been accompanied by the advent of high-speed programmable networking substrates which have drastically increased the ability to dynamically change packet-processing functionality. The switching chips for these substrates \cite{tofino, xpliant, flexpipe, rmt, drmt} have demonstrated that relative to fixed-function chips, a certain level of programmability can be achieved without compromising performance within the data plane. Along with these switching chips, high-level domain-specific languages for data plane programming such as P4 \cite{p4-14}, Domino \cite{domino}, POF \cite{pof}, and Lyra \cite{lyra} have emerged to configure packet-processing behavior. With these advances, programmable switches have been shown to have a plethora of use cases. These uses dynamically implementing new protocols such as VXLAN \cite {vxlan}, running network functions such as firewalls or load balancers \cite{gallium}, and implementing centralized lock management \cite{netlock}.

Today, most programmable switching chips feature a pipeline of stages that perform packet-processing computations. However, building compilers for these chips remains challenging. Unfortunately, programmers bear the weight of these consequences as they rely on compiler heuristics to adequately map their programs to machine code and an incorrect mapping could result in a binary with erroneous behavior. While the testing and development of traditional compilers has never been easy, the issue is exacerbated for compilers targeting switches. We make the following observations: 
\raggedbottom
\begin{CompactEnumerate} 
\item Switching chips have \textit{restraining budgets} of hardware resources such as pipeline stages and arithmetic logic units (ALUs).
\item Due to the \textit{feed-forward} design of switch pipelines where packets flow from an earlier stage to a later one but not in reverse, computations must be placed into stages while respecting dependencies between computations.
\item Programmable pipelines have an \textit{all-or-nothing} nature, meaning that a program either runs at line-rate if it can fit within a pipeline's resources or it is rejected by the compiler.
\end{CompactEnumerate} 
Furthermore, severe damage can result from bugs whose effects can permeate across an entire network causing issues such as security vulnerabilities if ACLs aren't correctly implemented, heightening the importance of validating compiler correctness.

 Thus, due to limited switch hardware resources as well as the feed-forward pipeline architecture, the mapping of programs to the switch computational model is challenging. This along with switches' all-or-nothing nature results in compilers having to frequently reject high-level programs. This is because compilers are unable to map programs to the underlying switching chip \cite{p4-studio, domino-compiler}, even if a mapping exists. To better illustrate the severity of this problem, consider the case where there exist two seemingly different, yet semantically equivalent programs. It is possible for these switch programs to have two different compilation results --- one can succeed in being compiled while the other fails and gets rejected. For general-purpose CPUs, if the high-level implementation of a program is not resource-efficient, it can still be executed regardless of the aggressiveness of the compiler optimizations at the costs of longer runtime and/or more memory usage. Compiler developers for general-purpose CPUs also don’t have to grapple with the resource-constrained pipeline model of programmable switches. Additionally, compiler developers for general-purpose CPUs have access to a plethora of robust testing and bug-finding tools and techniques (\eg \cite{csmith, spe}) that aren't available to switch compiler developers. In summary, the design and implementation of compilers for packet-processing is much more challenging than for traditional compilers. A few questions are posed because of these difficulties: how can we ease the tedious development process for switch compiler developers? How can we check the correctness of switch compiler mappings to the switch pipeline instruction set? In this paper, we address these concerns.

We present \textbf{Druzhba}, a switch pipeline hardware simulator for testing compilers targeting high-speed programmable switches. To test their compilers, developers target Druzhba as a compiler backend. Druzhba then simulates compiler-generated machine code programs so that compiler developers can observe whether their programs exhibit the expected packet-processing behavior or not. If unexpected behavior is observed, compiler developers conclude that the compiler mapping to the switch pipeline instruction set was erroneous. In implementing Druhba, we model the low-level hardware primitives of the RMT (Reconfigurable Match Tables) \cite{rmt} architecture. We do this by enabling compiler developers to specify the low-level hardware implementation details of the switching chips that they are programming their compilers to target. We are also in the process of implementing simulation for a network processor-based model, dRMT (Disaggregated Reconfigurable Match Tables) \cite{drmt}. Druzhba's source code can be found at \url{https://github.com/chipmunk-project/druzhba-simulator}.

\raggedbottom
\section{Switching Chip Architecture}

In this section we discuss the high-speed packet forwarding performed by switches. We also delve into the emergence of the RMT programmable switch architecture that we model. We further describe our RMT pipeline instruction set modeling methodology.

\begin{figure*}[!t]
\includegraphics[width=\textwidth]{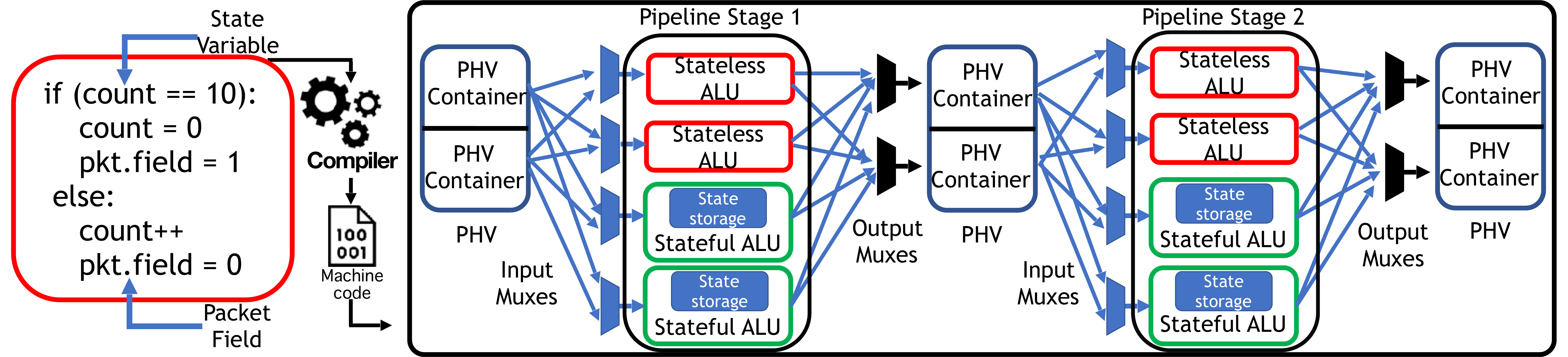}
\vspace{-0.3in}
\caption{The left side shows a high level program (\eg Domino, P4) given to the compiler. It is mapped to the Druzhba RMT machine model on the right, which features a pipeline with depth and width of 2 and PHV length of 2. The pipeline stages' connections between PHVs to ALU inputs and ALU outputs to PHVs are also shown.}
\vspace{-0.15in}
\end{figure*}

\subsection{Overview}
Switches perform high-speed packet forwarding which first involves a parser to extract packet fields from an incoming bytestream. Second, they operate on packets using match+action tables. These tables are allocated using local pipeline stage memory and map matches on packet header fields to actions that perform computations on packet header fields, metadata, and switch state. Examples of actions include mutating a state variable, dropping a packet, or decrementing a packet’s TTL. 

\Para{Motivation for switch chips. }\textcolor{black}{CPUs and network processors initially come to mind as ideal candidates for these processing requirements but they do not perform at high speeds. CPUs have a general-purpose instruction set and use external memory and thus aren't specialized to perform fast packet-processing. Many network processors contain computing clusters with configurable units for packet processing. However, these architectures have variable performance, don't guarantee line rate, and can have degraded performance when memory is shared between different processors. On the other hand, switching chips can operate at two orders of magnitude faster than many CPUs and one order of magnitude faster than many network processors. }

\textcolor{black}{One prominent fixed-function switching chip design was the Multiple Match Tables (MMT) model which consists of two pipelines, called the ingress and egress pipelines, that are separated by switching fabric which determines the connections between the input and output ports. Each pipeline is comprised of a series of pipeline stages stages with each stage containing local memory to be used for match+action tables. Due to the performance requirements of line-rate forwarding, fixed-function MMT chips limit the freedom in switch reconfiguration which is problematic for implementing new header fields for matching and actions for tasks such as tunneling, queue management, and traffic engineering. }

\Para{Programmable pipelines.} 
 The RMT architecture has risen as a popular alternative due to its increased flexibility. RMT also contains pipelines of match+action tables but goes further in allowing reconfigurability of the switching chip data plane. The first contribution is that the parser is programmable, enabling new header types and fields to be defined without being restricted to pre-defined ones. Second, the size and number of match tables within the switch can be reconfigured. Third, new actions that haven't been pre-defined can be created. Lastly, more control is given in allowing packets to be placed in specific queues. The design of RMT's match+action tables reduces wasteful resource consumption and allows for the ability to meet different algorithmic requirements. On the other hand, in the MMT model, new hardware often needs to be constructed for a specific configuration that a current switch does not support.

\subsection {Compilation to Switch Pipelines} 
Along with the increased freedom in programmability, compilers are responsible for ensuring that high-level programs are mapped to switch hardware primitives. Within the hardware, the parser generates packet header vectors (PHVs) that are vectors of containers each holding a packet or metadata field; metadata is data associated with each packet. Metadata fields include the number of bytes in the packet or the ingress port on which the packet arrived. Action units are implemented using configurable digital circuits which comprise arithmetic logic units (ALUs) and memories. ALUs perform computations and are either stateful or stateless; stateful ALUs can read and write to their switch state values while stateless ALUs solely operate on PHVs. Switch state is data that is stored locally within an ALU and any modification made to state must be visible to the next PHV that the ALU executes on. State is local to stateful ALUs and isn't shared across multiple ALUs to prevent non-deterministic performance from memory contention. Compilers translate programs to machine code using the instruction set of the underlying switching chip to (1) determine which header fields for a parser to match on and place into PHVs, (2) implement the tables and ALUs, and (3) generate the connections between ALUs and PHVs. Figure 1 shows the compilation process of taking a high-level packet-processing program and converting it to switch pipeline machine code. The machine code is then used to program the hardware primitives of our Druzhba machine model.

\subsection{RMT Instruction Set Modeling} 
\label{sec:instrustion-set-model}

Druzhba doesn't directly represent the match+action tables, but models their underlying hardware primitives. First, instead of modeling packets directly, we model PHVs to better capture the low-level architectural details. Second, we use ALUs to represent the switch action units. Third, we use input and output multiplexers to illustrate the connections between PHVs and ALUs. Druzhba accurately models the ALUs within a physical switching chip (e.g., \cite{tofino}). At the moment, we do not model parsing, matching, and other switching chip functionalities.

 ALU behavior is determined by opcodes that specify the type of operations to perform and immediate operands that are unsigned integer constants. PHV container values are fed into an ALU through input multiplexers with each multiplexer corresponding to an ALU operand. Once the input multiplexers have forwarded the operands to their respective ALUs, the ALUs execute and state variables are written to as needed. Each output multiplexer receives the output value from every each ALU and selects one to write to its allocated PHV container. Figure 1 shows an in-depth view of our model by illustrating Druzhba's feed-forward pipeline structure and the multiplexers that connect the PHVs and ALUs.

\section{Design and Implementation}

Our Druzhba pipeline simulation consists of (1) our pipeline generator, dgen, and (2) our simulation component, dsim, which performs the packet-processing behavior specified by dgen's generated pipeline on incoming packets. In this section, we delve into these details as well as how we employ optimizations to simplify the pipeline code and reduce dsim simulation runtime. Druzhba is written entirely in Rust.

\subsection{Hardware Specification}

We express our pipeline model by allowing dgen to take specifications of the hardware and convert them into an executable Rust program of the pipeline given (1) the depth and width of the pipeline (\ie number of stages and number of ALUs per stage), (2) a high-level representation of the ALU structure, and (3) machine code to determine the switch's behavior. The machine code programmatically defines the behavior of the multiplexers and ALUs. The pipeline model that is generated by dgen is the design that will be simulated for compiler testing. Each pipeline configuration is defined by the width and length of the pipeline in addition to the high-level representation of the ALUs. This customizability thus effectively allows Druzhba to act as a family of simulators, one for each possible pipeline configuration.

\Para{Expressing ALU functionality.} We express the capabilities of ALUs via our ALU domain-specific language (DSL). Our ALU DSL allows us to specify the input packet field operands and state variables, whether the ALU is stateful or stateless, and the immediates and opcodes that determine the ALU's computations. The input packet fields come from the PHV container values. The ALU DSL supports unary and binary ALU operations as well as additional multiplexers; binary operations can use either arithmetic, relational, or logical operators. Figure 2 shows our ALU DSL grammar. We have written 5 stateless ALUs and 6 stateful ALUs in our ALU DSL that model atoms in Banzai \cite{banzai}, a switch pipeline simulator for Domino. Atoms are Banzai's natively supported atomic units of packet-processing.

Stateless ALUs take in an opcode, a set of PHV container values, and an immediate operand as input. These ALUs perform an arithmetic operation on its operands which is determined by the opcode. Using our ALU DSL, these semantics can be captured using an if statement. Stateful ALUs also take in a set of PHV containers as input as well as state variables. Stateful ALUs are much more constrained in their computational abilities since they must be able to read and write to state while still maintaining the switch's throughput of 1 packet per clock cycle. Because of these difficulties, stateful ALUs are more complex and contain more configurable operations. Figure 3 shows one of our stateful ALUs that models Banzai's If Else Raw atom \cite{banzai}.

\Para{Machine code for switch primitives.}
To interface with Druzhba and to configure pipeline behavior, a compiler-generated machine code program consisting of a list of string and integer pairs is provided. The machine code pairs correspond to pipeline hardware primitive operations and are each identified by a unique name. The integer value in each pair determines the behavior of that operation. In Figure 3, the expressions $C()$, $arith\_op()$, $rel\_op()$, and $Opt$() are configurable ALU operations whose behaviors are determined by machine code values such as opcodes and immediates. For instance, an ALU relational operation, $rel\_op(e_{1}, e_{2})$, uses an opcode to determine which relational operator ($!$$=$, $<$, $>$, $==$) to perform on expressions $e_{1}$ and $e_{2}$. Our machine code also allows for determining the connections between PHVs to ALU inputs and ALU outputs to PHVs through specifying the behavior of the input and output multiplexers. For instance, a 3-to-1 input multiplexer uses its multiplexer control setting within the machine code to determine which of its 3 PHV container values to send to the connected ALU.

\begin{figure}
\newcommand{\sep}{~|~}
\begin{scriptsize}
\begin{eqnarray*}
l \in \text{literals} \quad v \in \text{variables} \quad &rel\_ op& \in \text{relational operators} \\
\quad &log\_ op& \in \text{logical operators}
\\
\quad &arith\_ op& \in \text{arithmetic operators} 
\\
\quad &un\_ op& \in \text{unary operators} 
\\
t \in \text{ALU type declaration} &::=& \text{\tt stateful} ~ | ~ \text{\tt stateless}\\
D \in \text{packet field declarations} &::=& \text{set of variables $v$} \\
K \in \text{Hole variable declarations} &::=&   \text{set of variables $v$}\\
SV \in \text{state variable declarations} &::=& \text{set of variables $v$} \\
e \in \text{expressions} &::=& l \sep v \sep e~arith\_ op~e  \sep e~rel\_ op~e \\
& & \sep un\_ op~e \sep  e~log\_op~e \sep {\tt Mux}(e, e, \ldots) \\ 
s \in \text{statements} &::=& e = e \sep {\tt if}~(e)~\{ s \} \sep {\tt if}~(e)~\{ s \}~{\tt else}~ \{ s \} \\
& & \sep s~;~s \sep {\tt return}~(e)  \\
%
%
%
p \in \text{ ALU specification} &::=& t; D ; K; SV; s
\end{eqnarray*}
\end{scriptsize}
\vspace{-0.3in}
\caption{Overview of the ALU DSL grammar. Operators include  logical ($\&\&$, $||$), relational ($!$$=$, $<$, $>$, $==$), arithmetic ($+$, $-$, $*$, $/$), and unary ($-$). Hole variable declarations include additional machine code values such as opcodes and immediate operands.}
\vspace{-0.15in}
\label{fig:grammar}
\end{figure}


\subsection {Compiler Testing Workflow} 
 \Para{Pipeline generation.} dgen uses the hardware specification and generates a Rust program that can be used to simulate a pipeline with that hardware specification. This generated program represents the scaffolding of the specified pipeline as well as the ALUs within it. Abstract Syntax Trees (ASTs) are generated from the given ALU files. As the ASTs are traversed, corresponding Rust code for the ALUs is generated. A Rust function is created for each ALU and subsequent helper functions are created for multiplexers and ALU operations. Each multiplexer and ALU operation function is given additional parameters for machine code values. This process is repeated for every stateful and stateless ALU within the switch pipeline. Once these ALU functions in addition to their corresponding helper functions are generated, additional code is generated to specify the overall structure of the pipeline. This code groups the ALU functions together by stage and connects the PHVs, multiplexers, and ALUs together. This code ensures that the input and output multiplexers as well as the ALUs are executed in the proper order within the pipeline. Further, it utilizes a hash table of machine code pairs and passes on the machine code values to the proper hardware units. For instance, it will give input multiplexer functions their proper machine code values needed to determine which operands to forward to their allocated ALUs. dgen's generated Rust program is written into a Rust file that we refer to as the pipeline description.

\begin{figure}[t!]
    \centering
    \includegraphics[width = \columnwidth]{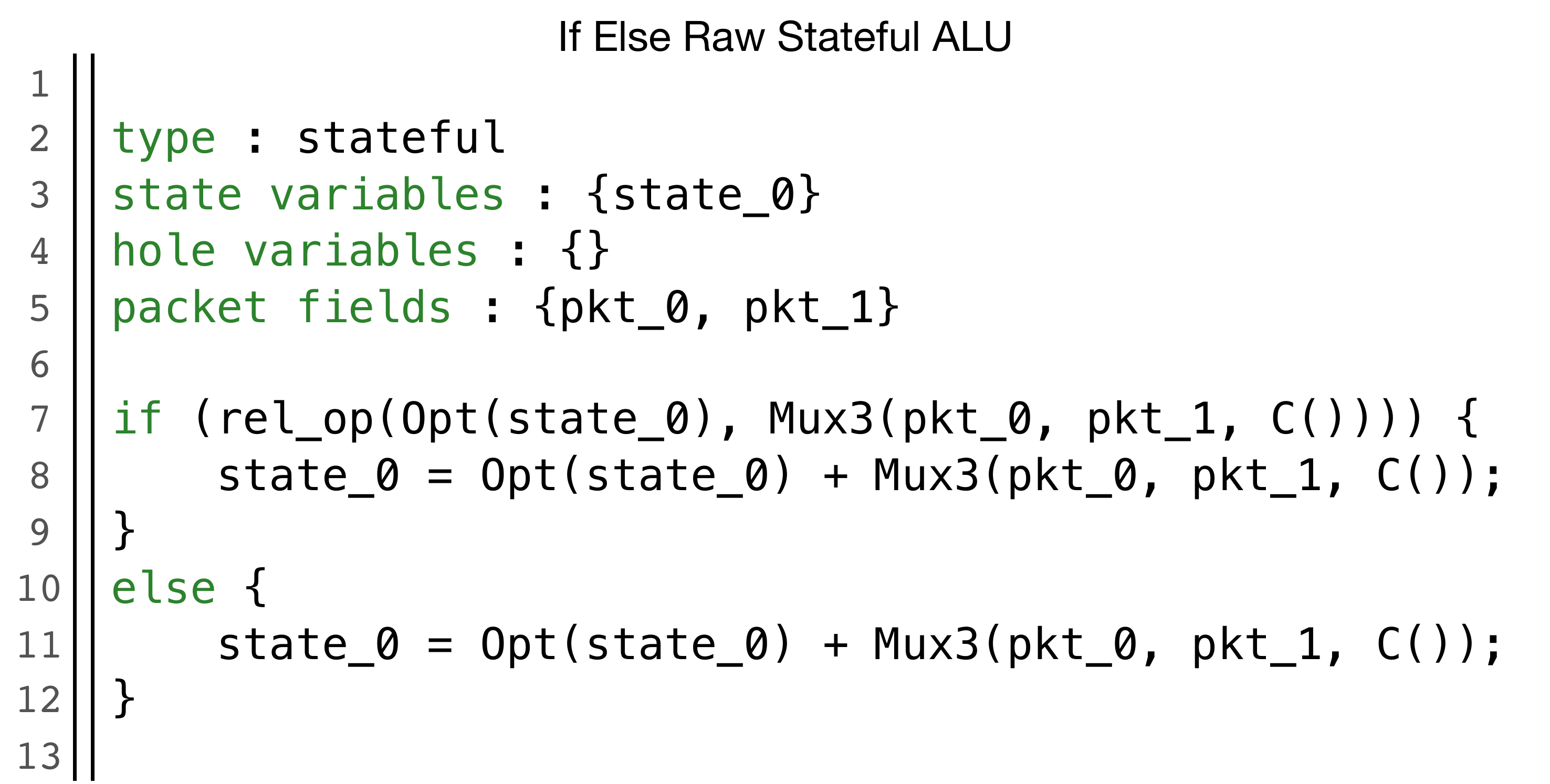}
    \vspace{-0.3in}
    \caption{If Else Raw Banzai \cite{banzai} atom written in the ALU DSL. C() denotes a constant and Opt() denotes a 2-to-1 multiplexer that either returns 0 or its argument.}
   \label{fig:stateful_alu}
   \vspace{-0.15in}
\end{figure}

\Para{Pipeline simulation.} After dgen's pipeline description is complete, it is compiled with dsim. Prior to the simulation of the architecture specified by the pipeline description, state variables are either initialized with random integer values or they are specified by the user. At every simulation tick, a PHV created by the traffic generator enters the pipeline and is executed by the first pipeline stage and PHVs in subsequent stages are sent to their next respective stages. We refer to the PHVs generated by the traffic generator along with the recorded state variable values at the time of each PHV entering the pipeline as the \textit{input trace}.  Following each simulated tick, an output PHV is generated. We call the modified PHVs along with the recorded state variable values at the time of each PHV leaving the pipeline the \textit{output trace}. dsim displays to the user both the input and output traces. 

Druzhba tests a compiler-generated program by fuzzing the generated pipeline model. This is done using the random PHVs generated by the traffic generator and checking the correctness of the output trace. When generating the PHVs, each PHV container was initialized with an integer in [0, 10000] --- it is straightforward to extend or restrain this bound if needed. The user then writes a high-level specification capturing the intended algorithmic behavior on both PHVs and state values and recording both the input and output traces. To write this specification, compiler developers convert the initial program given to the compiler into Rust code. The input trace is then given to the specification which generates its own output trace. A compiler-generated machine code program is said to be correct if the output trace generated by the specification is equivalent to the output trace generated by dsim. In other words, the program is correct if the output PHVs at every tick and the values of the state variables at those ticks are equivalent to the corresponding output PHV container and state variable values produced by the specification program. Conversely, a machine code program is said to have failed if there exists an input PHV and state variable configuration such that the simulation yields a different output from the specification program. Figure 4 shows this compiler testing process. We use Rust assertions to determine whether output traces containing the output PHVs and modified state variables from dsim and the specification match or not.

\subsection {Optimizations} 

Compiler developers need to quickly test their compilers on many different programs at once. To gauge overall compiler correctness, they need to test mappings from a wide distribution of varying programs. However, these program simulation times can become cumbersome during compiler development. This is especially true when frequent changes are made to the compiler codebase or when the compiler tester wants to increase the number of input PHVs to enable a larger PHV search space to be explored. We thus seek to reduce dsim's simulation time.

\begin{figure}[t!]
    \centering
    \includegraphics[width = \columnwidth]{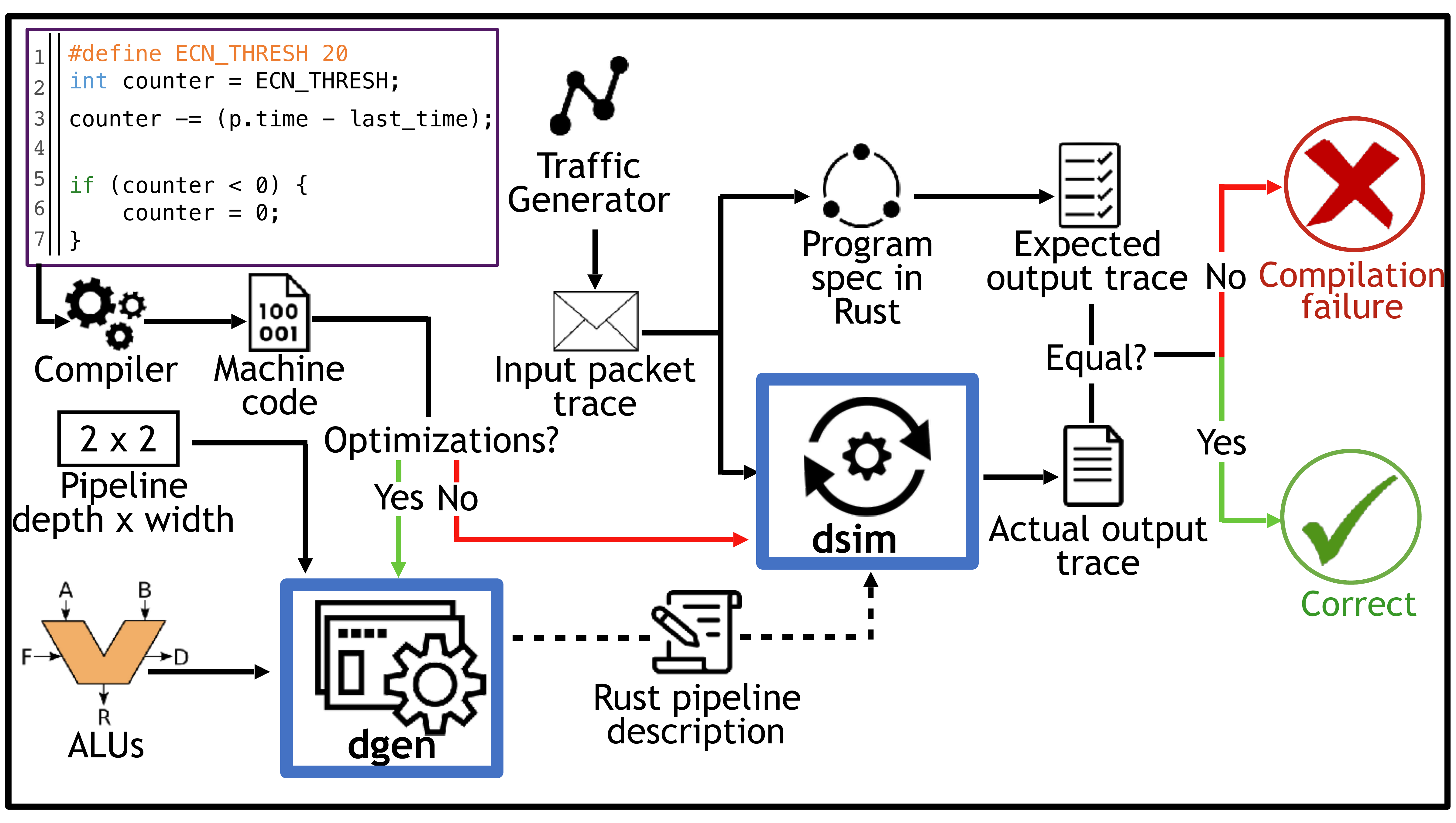}
    \vspace{-0.3in}
    \caption{Compiler testing workflow.}
   \label{}
   \vspace{-0.15in}
\end{figure}

\Para{Sparse conditional constant propagation.} dsim initially took machine code in as input instead of dgen which caused the pipeline description functions to treat the machine code as variables that are passed as arguments during runtime. This allowed machine code to be swapped between simulations without rerunning dgen and recompiling dsim. In beginning optimizations, we give the machine code as input to dgen and note that (1) providing the machine code pairs during pipeline generation enables a global static mapping of names to values which is unchanged for the duration of the simulation and (2) the functions in our pipeline description use if statements to check these values. These observations allow us to use sparse conditional constant (SCC) propagation \cite{scc_prop}, which involves constant propagation followed by the evaluation of branch conditions, which in turn allows it to remove entire code paths that will never be
evaluated. We do this by replacing machine code variable occurrences with their corresponding integer values. Then we use constant folding by evaluating constant expressions which allows us to determine the results of conditional statements. This results in dead code elimination from unused control paths and solely emitting single simplified expressions in place of the previous function bodies.

For instance, consider an arithmetic operation function that adds its operands if its machine code opcode is 0 and subtracts otherwise. During optimization, the if statement that checks the opcode is removed and solely replaced with either the addition or subtraction expression. Large opcode values can cause function behavior to branch in many different ways. This initially required numerous conditional expressions to check against every possible value case but now these computations are not performed during simulation.

\section{Evaluation}
In this section we describe our experience in testing a compiler for switch pipelines. We also evaluate the simulation performance of Druzhba on several benchmark programs.

\subsection {Case Study}
Druzhba tested Chipmunk \cite{chipmunk}, a retargetable compiler for packet-processing pipelines. Chipmunk compiles programs written in Domino \cite{domino}, a high-level language for programmable switches with C-like syntax, to machine code using program synthesis \cite{synthesis}. At this time, we have only tested Chipmunk. Chipmunk's usage of program synthesis was used to easily interface with Druzhba's instruction set. Testing other existing compilers would require further manual work in enabling them to generate the proper machine code to interface with Druzhba's instruction set. To set up our testing of Chipmunk, we first created multiple Domino programs and generated corresponding machine code for each one. We then took each of the Domino programs that were initially given to Chipmunk and converted them to Rust; these programs served as our specification programs. During the testing process, for each program we fuzzed both the switch pipeline model and the corresponding program specification. We checked that the state variables and output PHVs were equivalent to the state variables and output PHVs from the specification program. These tests validated the accuracy of Chipmunk's code generation for the programs tested. Table 1 shows 12 Domino programs that were compiled to machine code by Chipmunk and later converted to Rust. 

Over 120 Chipmunk machine code programs were determined to be correct after fuzz testing using Druzhba. These programs were generated by taking the 12 benchmarks and mutating them in semantic-preserving ways to create more test cases for the Chipmunk compiler. This shows how
Druzhba can be used as an aid during compiler development to improve the correctness of compilers.

\subsection{Benchmarks}
We execute our benchmarks by taking 12 compiler-generated programs from Chipmunk and measuring the amount of time it took to perform unoptimized and optimized simulations for 50000 PHVs for each one using Rust's supported benchmark tests. These programs are listed in Table 1. All experiments were performed with our 28-core 56-hyperthread 64-GB RAM machine (Intel Xeon Gold 6132). For each program, we measured the amount of time it took to run (1) the unoptimized simulation and (2) the optimized pipeline simulation using SCC propagation. The program complexity and number of PHV containers the program uses dictate the pipeline dimensions needed to implement the intended algorithmic behavior.

Generally, programs in Table 1 that showed the most significant improvements due to our optimizations were the ones with the highest number of pipeline depths and widths, \eg stateful firewall, flowlets, and learn filter. Since the pipeline code generated is commensurate with pipeline size, unoptimized runtime was much higher and the optimizations affected a greater portion of code for larger pipeline simulations. We also looked at how ALU program complexity affected simulation times since every ALU is executed for every PHV that traverses the pipeline. Some ALU implementations were much more terse and consisted of much fewer arithmetic, relational, and logical operations than others. But we found that the ALUs' program complexities had little impact on performance.

\begin{table}[!t]
\begin{center}
\footnotesize
\centering
    \begin{tabular}{|p{2.3cm}|p{0.7cm}|p{1cm}|p{1.4cm}|p{1.1cm}|}
         \hline
         \textbf{Program} & \textbf{Depth, width} & \textbf{ALU name} &  \textbf{Unoptimized (ms)} & \textbf{Optimized (ms)}     \\
         \hline
         BLUE (decrease) \cite{blue} & 4,2 & sub & 986 & 576 \\
         \hline
         BLUE (increase) \cite{blue} & 4,2 & pair & 1,268 & 724 \\
         \hline
         Sampling \cite{domino} & 2,1 & if else raw & 234 & 167 \\
         \hline
         Marple new flow \cite{marple} & 2,2 & pred raw & 404 & 215 \\
         \hline
         Marple TCP NMO \cite{marple} & 3,2 & pred raw & 729 & 481 \\
         \hline
         SNAP heavy hitter \cite{snap} & 1,1 & pair & 143 & 103 \\
         \hline
         Stateful firewall \cite{snap} & 4,5 & pred raw & 1,549 & 703 \\
         \hline
         Flowlets \cite{flowlet} & 4,5 & pred raw & 1,771 & 983 \\
         \hline
         Learn filter \cite{domino} & 3,5 & raw & 1,911 & 1,162 \\
         \hline
         RCP \cite{rcp}& 3,3 & pred raw & 1,261 & 793 \\
         \hline
         CONGA \cite{conga} & 1,5 & pair & 393 & 206 \\
         \hline
         Spam detection \cite{snap} & 1,1 & pair & 145 & 103 \\
         \hline
    \end{tabular}'
    \vspace{0.04in}
    \caption{Simulation runtimes with and without optimizations. ALU names refer to Banzai \cite{banzai} atoms.}
    \label{tab:results}
    \vspace{-0.35in}
\end{center}
\end{table}

\section {Related Work}
\label{sec:related_works}
Network simulation tools have long been used. Platforms such as Mininet \cite{mininet} and CrystalNet \cite{crystalnet} are primarily concerned with emulating a network of data communications devices. PFPSim \cite{pfpsim} is more similar to Druzhba and models the architecture of programmable switches and simulates match+action operations using P4 programs. NS4 \cite{ns4} is also a switch simulator but goes a step further from PFPSim by allowing emulation of entire P4-enabled networks. While PFPSim and NS4 are useful for debugging and studying algorithmic impact at the networking level, Druzhba is a platform that takes a different approach by modeling the low-level details of the switch pipeline instruction set to test compilers that target switch pipelines. \textcolor{black}{Program testers interface with PFPSim and NS4 using high-level P4 programs while compiler developers interface with Druzhba using switch pipeline machine code after their programs have been compiled. }

Though Banzai \cite{banzai} is a switch simulator that serves as a compiler target for Domino, it does not model the switch architecture at the same low-level detail as Druzhba. This low-level modeling allows Druzhba to simulate machine code, which Banzai cannot do. To better illustrate their differences, consider the case of simulating a compiled Domino program using Banzai. The Domino compiler generates code at a relatively high level of abstraction --- the source program is sliced into codelets that specify the high-level behavior to be performed by the atoms\footnote{Domino refers to these ALUs as atoms.}. Then the Domino compiler verifies that a mapping from codelets to atoms exists. But Banzai only performs direct execution of these codelets within the context of the pipeline rather than executing atoms themselves. On the other hand, when using Druzhba, the program is compiled down to machine code. During simulation, the ALUs are executed and perform the behavior specified by the machine code. Thus, because Druzhba more closely resembles the low-level hardware details of the switching chip, there is a higher level of assurance that compiler mappings are correct. Because Druzhba provides a more comprehensive end-to-end check of compiler correctness, bugs that are deeper within the compilation process can be found (\eg backend code generation bugs).

\section{Future Work}
We recognize that there is still room for additional amelioration and development. First, our RMT simulation can be further enhanced by modeling other switch details such as buffers, tables, and parsing and the interface can be improved to ease the difficulty of Druzhba code generation. Second, our dRMT simulation that simulates high-level P4-14 programs isn't comprehensive; many details such as packet field length aren't thoroughly simulated. Further, we look to use program verification by allowing support for a high-level specification that contains the pipeline's intended algorithmic behavior as well as PHV and state value constraints. This specification and the pipeline description will be transformed into SMT formulae so that equivalence will be formally proven, instead of being checked by fuzz testing. We also look to adding a domain-specific time travel debugger \cite{time_travel_debugging} for Druzhba to further aid in bug finding. Lastly, we are looking into using Druzhba to evaluate the impact and effects of new hardware designs by modeling different instruction sets or by adding hardware support for multitenancy \cite{multitenancy}.

\section{Conclusion}
We presented Druzhba, a programmable switch simulator that performs low-level RMT instruction set modeling. We showed how Druzhba serves as a compiler target to test compilers for programmable switches. Druzhba has been useful in the testing of a program-synthesis-based compiler by simulating generated machine code programs. By simulating compiled machine code programs, Druzhba helped in determining whether the compiler had successfully mapped the high-level packet-processing programs to the low-level architectural details of our simulated pipeline instruction set. In the future, we expect that Druzhba can further aid in testing switch compilers for not only pipeline-based models, but also for other switching chip architectures.

\section*{Acknowledgements}
We are grateful to our shepherd, the anonymous CoNEXT reviewers, Suvinay Subramanian, and Tao Wang for their helpful and constructive comments on previous drafts of our paper. We thank Xiangyu Gao for his detailed answers to our Chipmunk questions.

\balance
\bibliographystyle{abbrv}
\bibliography{druzhba_conext}

\end{document}